\documentclass[graybox,psfig]{svmult}

\usepackage{mathptmx}       
\usepackage{helvet}         
\usepackage{courier}        
\usepackage{type1cm}        
\usepackage{makeidx}         
\usepackage{graphicx}        
\usepackage{wrapfig}
\usepackage{multicol}        
\usepackage[bottom]{footmisc}
\usepackage{amsmath}
\usepackage{subfigure}
\usepackage{url}

\makeindex             

\begin{document}

\title*{Self-organized criticality and adaptation in discrete dynamical networks}
\titlerunning{Self-organized critical networks} 
\author{Thimo Rohlf and Stefan Bornholdt}
\authorrunning{Rohlf and Bornholdt} 
\institute{Thimo Rohlf \at Max-Planck-Institute for Mathematics in the Sciences, 
Inselstrasse 22, D-04103 Leipzig, Germany \email{rohlf@mis.mpg.de}
\and Stefan Bornholdt \at University Bremen, Institute for Theoretical Physics, 
Otto-Hahn-Allee, 28359 Bremen, Germany \email{bornholdt@itp.uni-bremen.de}}
\maketitle

\abstract{It has been proposed that adaptation in complex systems is optimized 
at the critical boundary between ordered and disordered dynamical regimes. 
Here, we review models of evolving dynamical networks that lead to 
self-organization of network topology based on a local coupling between 
a dynamical order parameter and rewiring of network connectivity,
with convergence towards criticality in the limit of large network size $N$.
In particular, two adaptive schemes are discussed and compared 
in the context of Boolean Networks and Threshold Networks: 
1) Active nodes loose links, frozen nodes aquire new links, 
2) Nodes with correlated activity connect, de-correlated nodes disconnect. 
These simple local adaptive rules lead to co-evolution of network topology 
and -dynamics. Adaptive networks are strikingly different from random networks: 
They evolve inhomogeneous topologies and  
broad plateaus of homeostatic regulation, 
dynamical activity exhibits $1/f$ noise and attractor periods obey 
a scale-free distribution. The proposed co-evolutionary mechanism of 
topological self-organization is robust against noise and does not 
depend on the details of dynamical transition rules. Using finite-size 
scaling, it is shown that networks converge to a self-organized 
critical state in the thermodynamic limit. Finally, we discuss 
open questions and directions for future research, and outline 
possible applications of these models to adaptive systems in diverse areas.} 

\section{Introduction} 
\label{sec:intro} 
Many complex systems in nature, society and economics are organized as 
networks of many interacting units that collectively process information 
or the flow of matter and energy through the system; examples are gene 
regulatory networks, neural networks, food webs in ecology, species 
relationships in biological evolution, economic interaction and the internet. 
From an abstract point of view, one can distinguish {\em network structure}, 
i.e. the (typically directed) graph that describes the wiring of interactions 
between the nodes the network is composed of, and {\em network dynamics}, 
referring to certain state variables assigned to the nodes which can change 
in response to inputs or perturbations from other nodes. In the case of the 
genome, for example, dynamics of regulatory networks, as captured in changes 
of gene expression levels, results from repression and -activation of gene 
transcription controlled by regulatory inputs (transcription factors) 
from other genes \cite{Davidson2001}.

A main characteristic of all these systems is that they evolve in time, 
under the continuous pressure of adaptation to highly dynamic environments. 
Since network topology and dynamics {\em on} the network are typically 
tightly interrelated, this implies a co-evolutionary loop between a 
time-varying network wiring and adaptive changes in the nodes' dynamics. 
For example, there is evidence from the analysis of gene regulatory networks 
that interactions between genes can change in response to diverse stimuli 
\cite{Luscombe2004}, leading to changes in network toplogy that can be far 
greater than what is expected from random mutation. In the case of nervous 
systems, it is evident that self-organization and adaptation processes have 
to continue throughout the lifetime of a network, since learning is a major 
function of such networks. In this context, a major conceptual challenge 
lies in the fact that, in order to properly function as information 
processing systems, adaptive networks have to be, on the one hand, highly 
robust against {\em random} (or dys-functional) perturbations of wiring 
and dynamics (noise) \cite{Barkai1997,Wagner2000,Anirvan2002}, and, on the 
other hand, stay responsive to essential cues (information) from the 
environment that can change in time. While robustness would clearly favor 
highly ordered dynamics that is basically insensitive to any perturbation,
sensitivity and adaptive pressure tend to favor an ergodic sampling
of the accessible state space. The latter comes with the risk of leading 
network evolution into regimes of chaotic dynamics with large parameter
ranges where network dynamics is not easily controlled \cite{Molgedey1992}.

Two interesting and interrelated questions arise:
First, is there a critical point, given by specific values of order parameters
that characterize network toplogy and -dynamics, where adaptive dynamics
with its delicate balance between robustness and flexibility is optimized?
Second, can we find simple, very general principles of network
self-organization from {\em local} co-evolutionary rules that couple network 
rewiring  and -dynamics such that the network globally evolves to this point?

In the inanimate world, phase transitions from ordered to disordered dynamics 
at critical values of a system parameter are found in several classes of 
many particle systems, as for example in ferromagnets, where the system 
can maintain spontaneous magnetization below the Curie temperature, while 
above this critical point disorder induced by thermal fluctuations wins.
Similar transitions form an organized to a disorganized state also have 
been observed in living systems, for example in enzyme kinetics 
\cite{Murray2002}, growth of bacterial populations \cite{Nicolis1977} and 
brain activity \cite{Kelso1992}. Most  biological networks are different 
in many regards from the many particle systems as considered in standard 
statistical mechanics. In particular, interactions between units are typically 
asymmetric and directed, such that a Hamiltonian (energy function) does not 
exist. Furthermore, to make global dynamical properties accessible despite 
the overall stunning degree of complexity found in these networks, a number 
of simplifying assumptions have to be made.

In this line, random Boolean networks (RBN) were proposed as simplified 
model of large gene regulatory networks \cite{Kauffman1969a,Thomas1973}.
In these models, each gene receives a constant number $K$ of regulatory 
inputs from other genes. Time is assumed to proceed in discrete steps. 
Each gene $i$ is either "on" or "off", corresponding to a binary state variable 
$\sigma_i \in \{0,1\}$, which can change at time $t$ according to a (fixed) 
Boolean function of its inputs at time $t-1$ (a more formal definition will 
be given in section \ref{subsec:rbndef}). RBNs can easily be generalized to 
a variable number of connections per node, and "biased" update rules 
\footnote{The bias is typically parameterized in terms of a stochastic 
control parameter $p$, which determines the probability that a particular 
input configuration generates the output "1".}. Despite its simple 
deterministic update rule, this model exhibits rich dynamical behavior. 
In particular, RBNs exhibit an order-disorder phase transition when each
unit has on average two inputs from other nodes\footnote{This critical 
connectivity $K_c = 2$ refers to the simplest case, when all Boolean 
functions have equal probability to occur. For the case of biased update 
rules, this generalizes to $K_c = 1/(2p(1-p))$.} \cite{Derrida1986a}.

Combinatorial and statistical methods have provided quite detailed knowledge 
about properties of RBNs near criticality 
\cite{Derrida1986,Flyvbjerg1988,Kauffman1993,Bhattacharjya1996,Sole1995,Luque1996,Bhattacharjya1996a,Bastolla1998a,Bastolla1998,Luque2000b,Luque2000a,Albert2000,Andrecut2005,Kaufman2005,Drossel2005,Klemm2005,Kaufman2006,Correale2006,Correale2006b,Kesseli2006,Rohlf2007a}.
The second class of discrete dynamical networks that we will consider are 
Random Threshold Networks (RTN) with sparse asymmetric 
connections (for details, cf. section \ref{subsec:rtndef}). 
Networks of this kind were first studied as diluted, non-symmetric 
spin glasses \cite{Derrida1987c} and diluted, asymmetric neural networks 
\cite{Derrida1987,Kree1987}. 
For the study of topological questions in networks, a version  with discrete 
connections $c_{ij}=\pm1$ is convenient and will be considered here. 
It is a subset of Boolean networks with similar dynamical properties. 
Random realizations of these networks exhibit complex non-Hamiltonian dynamics 
including transients and limit cycles \cite{Kuerten1988b,Bastolla1996}. 
In particular, a phase transition is observed at a critical average 
connectivity $K_c$ with lengths of transients and attractors (limit cycles) 
diverging exponentially with system size for an average connectivity larger 
than $K_c$. A theoretical analysis is limited by the non-Hamiltonian 
character of the asymmetric interactions, such that standard tools of 
statistical mechanics do not apply \cite{Derrida1987}. 
However, combinatorial as well as numerical methods provide a quite 
detailed picture about their dynamical properties and correspondence 
with Boolean Networks  
\cite{Kuerten1988a,Kuerten1988b,Bastolla1996,Derrida1986,Derrida1986a,Derrida1987b,Flyvbjerg1988,Luque1996,Bastolla1998,Rohlf2002,Nakamura2004,Rohlf2007a,Rohlf2007c}.  

From the observation that complex dynamical behavior in these simple 
model systems is primarily found near criticality, Kauffman 
\cite{Kauffman1969a,Kauffman1993} and other researchers
\cite{Wooters1990,Langton1991} postulated that evolution should drive 
living systems to this "edge of chaos". Indeed, a number of parameters 
that are highly relevant for biological systems, as, for example, 
robustness \cite{Kauffman1993} and basin entropy \cite{Krawitz2007} of 
attractors (limit cycles), mutual information in the switching dynamics 
of nodes \cite{Luque2000b,Ribeiro2008} and information diversity in 
structure-dynamics relationships \cite{Nykter2008} are maximized near 
the order-disorder transition of RBNs, supporting the idea that this point 
provides unique properties for balancing the conflicting needs of 
robustness and adaptive flexibility. Today, experimental results provide 
strong support for the idea that many biological systems operate 
in a regime that shares relevant properties with criticality in random 
networks.
Indications for critical behavior were found, for example, in gene expression 
dynamics of several organisms \cite{Shmulevich2005,Ramoe2006,Nykter2008a} 
and  in neuronal networks in the brain \cite{Linkenkaer-Hansen2001,Beggs2008}. 
Since, in all these systems, there generally exists no central control 
that could continously adjust system parameters to poise dynamics at 
the critical state, we are {\em forced} to postulate that there are simple, 
{\em local} adaptive mechanisms present that are capable of driving 
{\em global} dynamics to a state of {\em self-organized criticality}.
Evolution towards self-organized criticality was established in a number
of non-equilibrium systems \cite{Bak1996}, namely, avalanche models 
with extremal dynamics \cite{Bak1987,Paczuski1996}, multi-agent models 
of financial markets \cite{Lux1999}, forest fires \cite{Malamud1998}  
and models of biological macroevolution \cite{Drossel1998}. 
Still, these approaches are limited in the sense that they consider
a fixed or at least pre-structured topology.

Network models of {\em evolving} topology, in general, have been studied 
with respect to critical properties earlier in other areas, e.g., in models 
of macro-evolution \cite{Sole1997}. Network evolution with a focus on gene 
regulation has been studied first for Boolean networks in \cite{Bornholdt1998} 
observing self-organization in network evolution, and for threshold networks 
in \cite{Bornholdt2000}. Combining the evolution of Boolean networks with 
game theoretical interactions is used for model networks in economics 
\cite{Paczuski2000}. 

Christensen et al. \cite{Christensen1998} introduced a static network 
with evolving topology of undirected links that explicitly evolves 
towards a critical connectivity in the largest cluster of the network. 
In particular they observed for a neighborhood-oriented rewiring rule that 
the connectivity of the largest cluster evolves towards the critical $K_c=2$ 
of a marginally connected network. However, in this model the core 
characteristics of adaptive networks, a co-evolution between dynamics 
and topology \cite{Gross2008}, is hard to establish, 
since the evolution rule, here chosen according to the Bak-Sneppen 
model of self-organized criticality \cite{Bak1993}, does not provide a 
direct coupling between rewiring of connections and an order parameter 
of the dynamics {\em on} the networks. 

Keeping the idea of local connectivity adaptations, a different line of 
research pursues models of adaptive co-evolutionary networks in the context 
of discrete dynamical networks, in particular based on RBNs and RTNs. 
The common principle in these models is the coupling of {\em local} 
rewiring events to approximate, local measurements of a dynamical order 
parameter. In the limit of large network sizes $N$, this principle leads to 
network evolution towards a {\em global} self-organized critical state.
Bornholdt and Rohlf \cite{Bornholdt2000a} introduced a topology-evolving 
rule based on the dynamical activity of nodes in RTNs: Active nodes, whose 
binary state changes in time, tend to lose links, while inactive (frozen) 
nodes, whose binary states are fixed, tend to gain new links. In  a recent 
extension \cite{Rohlf2007}, also adaptive changes of the nodes' activation 
thresholds were considered. A very similar co-evolutionary rule was applied to 
RBNs by Liu and Bassler \cite{Liu2006}; besides the case where only the 
rewired node is assigned a new Boolean function, they also consider 
"annealed" networks, where each node is assigned a new logical function in 
each evolutionary time step. Teuscher and Sanchez \cite{Teuscher2001} showed 
that this adaptive principle can also be applied to turing neural networks. 
Self-organized critical neural networks with stochastic dynamics and a 
rewiring rule based on dynamical correlations bettween nodes was studied by 
Bornholdt and R\"ohl \cite{Bornholdt2003}, observing robust self-organization
of both network toplogy and -dynamics. In the same context, Bertschinger et 
al.\ \cite{Bertschinger2005} studied a synaptic scaling rule leading to 
self-organized criticality in recurrent neural networks. A different adaptive 
scheme, based on a input-dependent disconnection rule and a minimal 
connectivity in RBNs, was studied by Luque et al. \cite{Luque2001}.  
A perturbation analysisindicates the emergence of self-organized critical 
behavior.

The remainder of this chapter is organized as follows: in section 
\ref{sec:rbndyn}, the dynamics of RBNs and RTNs are defined and basic 
dynamical and statistical properties of these systems are summarized. 
In particular, central order parameters that are relevant for the definition 
of adaptive algorithms will be introduced. In section \ref{sec:soc}, we 
will review different models of adaptive, discrete dynamical networks 
leading to evolution towards self-organized criticality that have been
established in this context so far, with a focus on activity- and 
correlation-based rewiring rules. 
Finally, section \ref{sec:summary} contains a summary and conclusions.

\section{Dynamics of Random Boolean Networks and Random Threshold Networks}
\label{sec:rbndyn}
In this section, we provide definitions for the two types of discrete 
dynamical networks under consderation, Random Boolean Networks and Random 
Threshold Networks. First, the underlying graph structure that connects 
dynamical units (automata) is defined, then dynamical update rules are 
provided. Further, basic dynamical properties of these systems are summarized.

\subsection{Underlying graph structure}
\label{subsec:graphs}
Concerning topology, discrete dynamical networks are described by 
{\em random directed graphs} $G(N,Z,g)$, where $N$ is the number of nodes, 
$Z$ the number of edges or links (arrows connecting nodes), and $g$ a function 
that describes the statistical distribution of the links between nodes. 
Arrows pointing {\em at} a node are considered as {\em inputs}, arrows 
pointing from this node {\em to} another node as {\em outputs}. If, for 
example, $Z$ links out of the $2N^2$ possible are assigned at random such 
that the average connectivity $\bar{K} := Z/N$ is fixed at a predefined value 
and $Z \ll 2N^2$ (sparse network), the resulting statistical distributions 
of the number $k$ of inputs and outputs follow a Poissonian \cite{ErdHos1960}:
\begin{equation}\label{Poissonconn_eq}
P(k) = \frac{\bar{K}^k}{k!}\exp{(-\bar{K})}.
\end{equation}
A schematic example of interaction graph structure is shown in the left 
panel of Fig. \ref{RBNscheme}.

\subsection{Random Boolean Networks}\label{subsec:rbndef}
A Random Boolean Network (RBN) is a discrete dynamical system composed of 
$N$ automata. Each automaton is a Boolean variable with two possible states: 
$\{0,1\}$, and the dynamics is such that
\begin{equation}
{\bf F}:\{0,1\}^N\mapsto \{0,1\}^N, 
\label{globalmap}
\end{equation} 
where ${\bf F}=(f_1,...,f_i,...,f_N)$, and each $f_i$ is represented by a 
look-up table of $K_i$ inputs randomly chosen from the set of $N$ automata. 
Initially, $K_i$ neighbors and a look-up table are assigned to each 
automaton at random.

\begin{figure}[htb]
\begin{center}
\includegraphics[scale=.6]{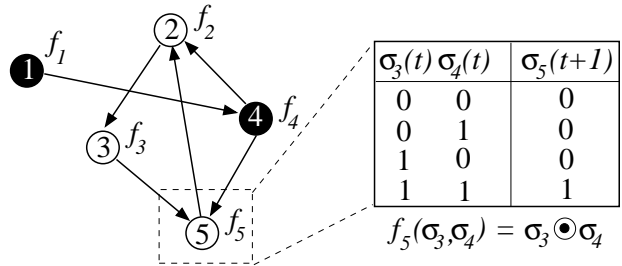}
\caption{Left panel: example of an interaction graph structure for a 
RBN of size $N=5$ with average connectivity $\bar{K} = 6/5$; 
$f_i$ are individual Boolean functions assigned to each node $i = 1,..,5$,
black circles mark $\sigma_i =1$, white circles $\sigma_i = 0$. 
Right panel: example of a Boolean update table assigned to a site 
(AND function of the site's inputs).}
\label{RBNscheme} 
\end{center}
\end{figure}

An automaton state $ \sigma_i(t) \in \{0,1\}$ is updated using its 
corresponding Boolean function:
\begin{equation}
\sigma_i(t+1) = f_i(\sigma_{i_1}(t),\sigma_{i_2}(t), ... ,
\sigma_{i_{K_i}}(t)).
\label{boolupdate_eq}
\end{equation}
We randomly initialize the states of the automata (initial condition 
of the RBN). The $N$ automata are updated synchronously using their 
corresponding Boolean functions, leading to a new system state 
$\Sigma:=(\sigma_1,...,\sigma_N)$:
\begin{equation}
\Sigma(t+1) = {\bf F}(\Sigma(t)).
\label{statevec_eq}
\end{equation}
The right panel of Fig. \ref{RBNscheme} provides an example of an 
individual update table assigned to a network site.
\begin{figure}[htb]
\begin{center}
\includegraphics[scale=.55]{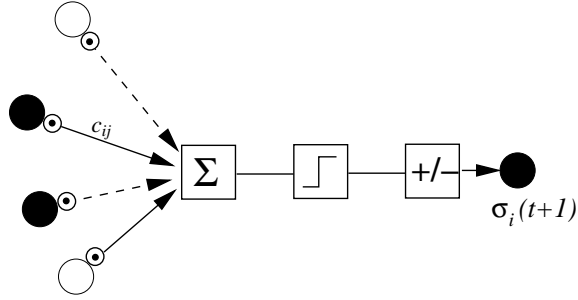}
\caption{Schematic sketch of a threshold dynamical unit: input states (circles 
on the left, black circles correspond to  a state $\sigma_j = +1$, white 
circles to $\sigma_j = -1$) are multiplied ($\odot$) with interaction weights 
$c_{ij}$ (lined arrows: $c_{ij} = +1$, dashed arrows:  $c_{ij} = -1$); these 
values are summed ($\Sigma$) and  added to a threshold. Finally, the output 
$\sigma_i(t+1)$ is determined by the sign ($+/-$) of the resulting signal.}
\label{RTNscheme} 
\end{center}
\end{figure}
\begin{figure}[htb]
\begin{center}
\includegraphics[scale=.55]{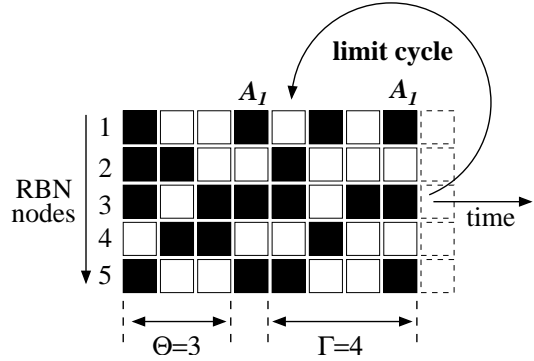}
\caption{Example of a dynamical trajectory for a $N=5$ Boolean network, time 
is running from left to right, network nodes are labeled from top to bottom. 
Black squares correspond to $\sigma_i = 1$, white squares to $\sigma_i = 0$. 
After a transient of $\Theta = 3$ system states $\Sigma$ the first state 
$A_1$ appears that repeats itself after $\Gamma = 4$ time steps, defining 
a periodic attractor (limit cycle) with period 4.}
\label{RBN_attrscheme}    
\end{center}
\end{figure}

\subsection{Random Threshold Networks}\label{subsec:rtndef}
A Random Threshold Network (RTN) consists of $N$ randomly interconnected 
binary sites (spins) with states $\sigma_i=\pm1$. For each site $i$, 
its state at time $t+1$ is a function of the inputs it receives from 
other spins at time $t$:
\begin{eqnarray} 
\sigma_i(t+1) = \mbox{sgn}\left(f_i(t)\right) \label{rtnupdate_eq}
\end{eqnarray}  
with 
\begin{eqnarray} 
f_i(t) = \sum_{j=1}^N c_{ij}\sigma_j(t) + h.  \label{rtnsum_eq}
\end{eqnarray}
The $N$ network sites are updated synchronously. In the following 
discussion the threshold parameter $h$ is set to zero. The interaction 
weights $c_{ij}$ take discrete values $c_{ij} = +1$ or $-1$ with equal
probability. If $i$ does not receive signals from $j$, one has $c_{ij} = 0$.

\subsection{Basic dynamical properties of RBNs and RTNs}\label{subsec:rbnrtnbaseprop}
Let us review a few aspects of the dynamics of Random Boolean Networks 
and Random Threshold Networks. In fact, they share most basic properties 
which is closely related to the fact that RTNs are a subset of RBNs. 

\subsubsection{Attractors and transients}\label{subsubsec:attr}
Update dynamics as defined in \ref{subsec:rbndef} and \ref{subsec:rtndef}, 
given the binary state $\sigma_i(t)$ of each node $i$ at time $t-1$, 
assigns a state vector $\Sigma(t)=(\sigma_1(t),...,\sigma_N(t))$ to 
the network at each discrete time step $t$. The path that $\Sigma(t)$ 
takes over time $t$ is a dynamical trajectory in the phase space of the 
system. Since the dynamics is deterministic and the phase space of the 
system is finite for finite $N$, all dynamical trajectories eventually 
become periodic. When we start dynamics from a random initial state, e.g. 
with each $\sigma_i(0)$, $i= 1 ... N$ set to $0$ or $1$ ($-1$ or $+1$ for 
RTN, respectively) independent from each other with equal probability 
$p = 1/2$, the trajectory will pass through $\Theta$ transient states 
before it starts to repeat itself, forming a limit cycles given by
\begin{equation}
\Sigma(t) = \sigma(t + \Gamma). \label{attr_def}
\end{equation}
The periodic part of the trajectory is the attractor of the dynamics, 
and the minimum $\Gamma \ge 1$ that satisfies Eq. (\ref{attr_def}) is 
the {\em period} of the attractor.

\subsubsection{Definition of average activity and average correlation}
\label{subsubsec:actdef}
Let us now define two {\em local} measures that characterize the typical 
dynamical behavior of a network site, and the dynamical coordination of 
pairs of sites. 

The \emph{average activity} $A(i)$ of a site $i$ is defined as the average 
over all states $\sigma_i(t)$ site $i$ takes in dynamical network evolution 
between two distinct points of time $T_1$ and $T_2$:
\begin{equation} 
A(i) = \frac{1}{T_2 - T_1+1}\sum_{t=T_1}^{T_2} \sigma_i(t)
\label{actdef_eq} 
\end{equation}
"Frozen" sites $i$ which do not change their states between $T_1$ and $T_2$ 
obviously have $|A(i)| = 1$ (or $|A(i)| = 0$, in the case of RBN), whereas 
sites that occasionally change their state have $0 \le |A(i)| < 1$.
\begin{figure}[b]
\begin{center}
\includegraphics[scale=0.9]{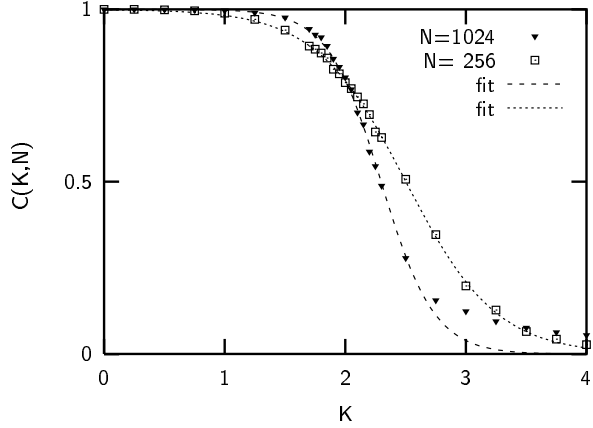}
\caption{The frozen component $C(K,N)$ of random threshold networks, as a 
function of the networks' average connectivities $K$. For both system 
sizes shown here ($N=256$ and $N=1024$) the data were measured along 
the dynamical attractor reached by the system, averaged over 1000 
random topologies for each value of $K$. One observes a transition 
around a value $K = K_0$ approaching $K_c = 2$ for large $N$. 
A sigmoid function fit is also shown. To avoid trapping in exponential 
divergence of attractor periods for $K > 2$, the simulations have 
been limited to $T_{max}=10000$. The mismatch of data and fit for 
$N=1024$, $K \ge 2.75$ is due to this numerical limitation.}
\label{frozcomp_fig}     
\end{center}
\end{figure}
\begin{figure}[b]
\begin{center}
\includegraphics[scale=0.9]{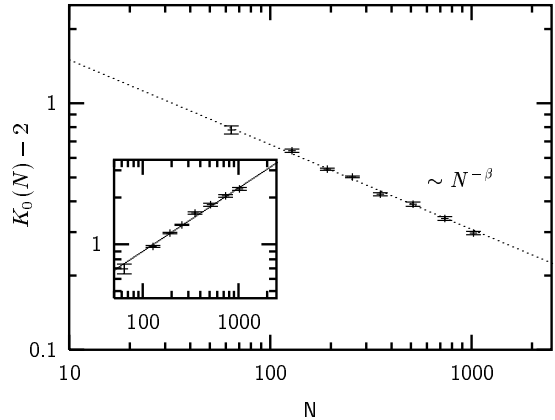}
\caption{The finite size scaling of the transition value $K_0$, obtained 
from sigmoidal fits as shown in Fig.\ \ref{frozcomp_fig}. $K_0$ approaches 
$K_c = 2$ with a scaling law $\sim N^{-\beta}$, $\beta=0.34 \pm 0.01$. 
The inset shows the scaling behavior of the parameter $\alpha(N)$; 
one finds $\alpha(N) \sim N^\gamma$, $\gamma = 0.41 \pm 0.014$.}
\label{froztrans_fig}      
\end{center}
\end{figure}
The \emph{average correlation} $\mbox{Corr}(i,j)$ of a pair $(i,j)$ of 
sites is defined as the average over the products $\sigma_i(t)\sigma_j(t)$ 
in dynamical network evolution between two distinct points of time $T_1$ 
and $T_2$:
\begin{equation} \mbox{Corr}(i,j) = \frac{1}{T_2 - T_1+1}\sum_{t=T_1}^{T_2} 
\sigma_i(t)\sigma_j(t) \label{corrdef_eq}
\end{equation}
If the dynamical activity of two sites $i$ and $j$ in RTN is (anti-)correlated, 
i.e.\ if $\sigma_i$ and $\sigma_j$ always have either the same or the opposite 
sign, one has $|\mbox{Corr}(i,j)| = 1$\footnote{In RBN, correlated pairs have 
$|\mbox{Corr}(i,j)| = 1$ and anti-correlated pairs $|\mbox{Corr}(i,j)| = 0$.}.
If the relationship between the signs of $\sigma_i$ and $\sigma_j$ 
occasionally changes, one has $0 \le |\mbox{Corr}(i,j)| < 1$. 

\subsubsection{Properties of $A(i)$ and $\mbox{Corr}(i,j)$ and their relation 
to criticality}
\label{subsubsec:actprop}
If we consider statistical ensembles of randomly generated networks with 
sparse wiring ($\bar{K} \ll N$), both $A(i)$ and $\mbox{Corr}(i,j)$ of 
RBNs and RTNs exhibit a second order phase transition at a critical average 
connectivity $K_c$ (averaged over the whole network ensemble). 
Below $K_c$, network nodes are typically frozen, above $K_c$, a finite 
fraction of nodes is active; this can be clearly appreciated from the 
behavior of the {\em frozen component} $C(\bar{K})$, defined as the 
fraction of nodes that do not change their state along the attractor. 
The average activity $A(i)$ of a frozen site $i$ thus obeys $|A(i)|=1$.
In the limit of large $N$, $C(K)$ undergoes a transition at $K_c$ 
vanishing for larger $K$. With respect to the average activity of a node, 
$C(K)$ equals the probability that a random site $i$ in the network has 
$|A(i)|=1$. Note that this is the quantity which is checked stochastically 
by the local rewiring rule that will be discussed in section 
\ref{subsubsec:actrewmodel}. The frozen component $C(K,N)$ is shown for 
random networks of two different system sizes $N$ in Fig. \ref{frozcomp_fig}.
One finds that $C(K,N)$ can be approximated by 
\begin{eqnarray} 
C(K,N) = \frac{1}{2} \{ 1+\tanh{[-\alpha(N)\cdot(K - K_0(N)\,)]} \}.  
\label{frozcompfit_eq}
\end{eqnarray}
This describes the transition of $C(K,N)$ at an average connectivity 
$K_0(N)$ which depends only on the system size $N$. One finds for the 
finite size scaling of $K_0(N)$ that
\begin{eqnarray} 
K_0(N) - 2 = a\cdot N^{-\beta} 
\label{frozcompfinite_eq}
\end{eqnarray}
with $a = 3.30 \pm 0.17$ and $\beta = 0.34 \pm 0.01 $ (see Fig.\ 4), 
whereas the parameter $\alpha$ scales with system size as 
\begin{eqnarray} 
\alpha(N) = b\cdot N^\gamma 
\end{eqnarray}
with $b= 0.14 \pm 0.016$ and $\gamma = 0.41 \pm 0.01$. This indicates 
that the transition of $C(K,N)$ exhibits a sharp decay near the critical 
connectivity $K_c$ when the thermodynamic limit  $N \rightarrow \infty$ 
is approached.

The number of frozen nodes is a decisive quantity for the evolution
of adaptive networks. If all nodes are frozen ($C = 1$), as it is typically
found for networks with very sparse $\bar{K}$, the network is basically 
irresponsive to signals from the environment and hence can neither process 
information nor adapt. If, on the other hand, $C$ vanishes, all nodes exhibit 
more or less chaotic switching behavior - dynamics becomes completely 
autonomous and hence again useless for information processing. A finite 
number of frozen nodes, as it is found near $K_c$, enables adaptive 
response to environmental signals by assignment of new, functional 
behavior to previously frozen nodes, and also makes sure that global 
network dynamics avoids the extremes of overly ordered and chaotic regimes. 
In the following section, we will discuss models of adaptive network
evolution by local dynamical rules that lead to emergence of self-organized
critical networks, i.e.\ networks that evolve to the "optimal" point
just at the phase transition from ordered to chaotic dynamics.

\section{Network self-organization from co-evolution of dynamics and topology}
\label{sec:soc}
In this section, we will discuss models of adaptive network self-organization 
in the context of discrete dynamical networks. The common principle that 
governs network evolution is a {\em co-evolution of dynamics and topology
from local dynamical rules}: An order parameter of network dynamics is 
estimated from local measurements (often averaged over a representative
number of dynamical update cycles, e.g., over one attractor period of a 
limit cycle the dynamics converged to, cf.\ section \ref{subsubsec:attr}). 
Based on the measured value of the order parameter, network connectivity 
and/or the switching behavior of nodes is adapted by local adaptive rules. 
Usually, there is a time scale separation between frequent dynamical 
updates and rare rewiring events. After a large number of adaptive cycles, 
evolution towards a {\em self-organized critical state} is observed. 

\subsection{Activity-dependent rewiring}\label{subsec:actdeprew}
\subsubsection{Motivation}\label{subsubsec:actdeptmotiv}
Living organisms process their information by dynamical systems at 
multiple levels, e.g.\ from gene regulatory networks at the cellular 
level, to neural networks in the central nervous system of 
multi-cellular organisms. As {\em complex adaptive systems}, organisms
have to deal with the conflicting needs of flexible response to changing
environmental cues, while maintaining a reasonable degree of stability in 
the dynamical networks that process this information. This led to the
idea that these systems may have evolved to the "edge of chaos" between 
ordered and disordered dynamical regimes \cite{Kauffman1993,Langton1991}. 
In the following, a simple evolutionary mechanism will be introduced
\cite{Bornholdt2000a}, based on a local coupling between a dynamical order 
parameter - the average activity of dynamical units (sites) in RTNs 
(Eq.\ \ref{actdef_eq}) - and a topological control parameter - the 
number of inputs a site receives from other units. In a nutshell, the 
adaptive rule can be summarized as {\em frozen nodes grow links, active 
nodes lose links}. This rule abstracts the need for both flexibility and 
stability of network dynamics. In a gene regulatory network, for example, 
a frozen gene cannot respond to different inputs it may receive, and hence 
is practically dysfunctional; the addition of a new regulatory input 
potentially assigns a new function to this gene. On the other hand, a very 
active gene will tend to show chaotic switching behavior and may lead to 
loss of stability in network dynamics - a reduction in input number reduces 
the probability of this undesirable behavior \cite{Rohlf2004a}. Similar 
demands for a local, homeostatic regulation of activity and connectivity 
can be expected in neural networks of the nervous system and are supported 
by experimental evidence \cite{Engert1999}.
\begin{figure}[htb]
\begin{center}
\includegraphics[scale=.65]{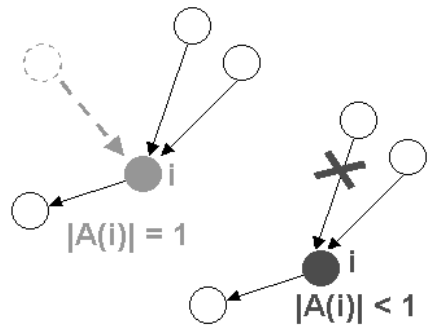}
\caption{The selective criterion leading to topological self-organization: 
A dynamically frozen site ($|A(i)|=1$) receives an additional regulatory input, 
an active site ($|A(i)|<1$) looses one of its inputs.}
\label{rewscheme}      
\end{center}
\end{figure}

\subsubsection{Model}
\label{subsubsec:actrewmodel}
Let us consider a Random Threshold Network of $N$ randomly interconnected 
binary elements as defined in section \ref{subsec:rtndef}. In the beginning, 
network topology is initialized as a directed, random graph with connectivity 
distributed according to a Poissonian with average connectivity $K_{ini}$ 
(cf. section \ref{subsec:graphs}), and $c_{ij} = +1$ or $c_{ij} = -1$ with 
equal probability for non-vanishing links. While network evolution is 
insensitive to $K_{ini}$ in general (as will be shown), we choose  
$0 < K_{ini} < 3$ in simulations to obtain reasonably fast convergence 
of the evolutionary dynamics. Network dynamics is iterated according to 
Eq.\ (\ref{rtnupdate_eq}) starting from a random initial state vector 
$\Sigma(0) = (\sigma_1(0),...,\sigma_N(0))$, with $\sigma_i =+1$  or 
$\sigma_i =-1$ with equal probability for each $i$. After $T$ iterations, 
the dynamical trajectory eventually reaches a periodic attractor 
(limit cycle or fixed point, compare section \ref{subsubsec:attr}). 
Then we apply the following local rewiring rule to a randomly selected 
node $i$ of the network: 
{\bf If node $i$ does not change its state during the attractor, 
it receives a new non-zero link $c_{ij}$ from a random node $j$.  
If it changes its state at least once during the attractor, 
it loses one of its non-zero links $c_{ij}$.} 
Iterating this process leads to a self-organization of the average 
connectivity of the network. The basic idea of this rewiring rule is sketched 
schematically in Fig.\ \ref{rewscheme}, a particular algorithmic realization 
is provided in Box 1.
\begin{svgraybox}
{\bf Box 1: Adaptive algorithm for activity-dependent rewiring}\\\\
This box gives an example of an adaptive algorithm that realizes the local 
rewiring rule "frozen nodes grow links, active nodes lose links" 
\cite{Bornholdt2000a}:
\begin{enumerate}
\item Choose a random network with an average connectivity $K_{ini}$. 
\item Choose a random initial state vector 
      $\Sigma(0)=(\sigma_1(0),...,\sigma_N(0))$. 
\item Calculate the new system states $\Sigma(t)$ according to eqn.\ (2), 
      using parallel update of the $N$ sites. 
\item Once a previous state reappears (a dynamical attractor with period 
      $\Gamma$ is reached) or otherwise after $T_{max}$ updates the simulation 
      is stopped. Then, a site $i$ is chosen at random and its average activity 
      $A(i)$ during the last $T = \Gamma$ time steps is determined (in case no 
      attractor is reached, $T=T_{max}/2$ is chosen).
\item If $|A(i)|=1$, $i$ receives a new link $c_{ij}$ from a site $j$ selected 
      at random, choosing $c_{ij}=+1$ or $-1$ with equal probability. If 
      $|A(i)|<1$, one of the existing non-zero links of site $i$ is set to zero. 
\item Finally, one non-zero entry of the connectivity-matrix is selected at 
      random and its sign reversed.
\item Go to step number 2 and iterate. 
\end{enumerate}
\end{svgraybox}
\begin{figure}[b]
\begin{center}
\includegraphics[scale=0.9]{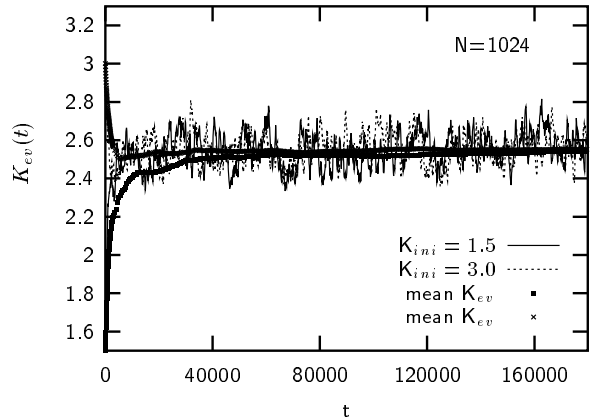}
\caption{Evolution of the average connectivity of threshold networks 
rewired according to the rules described in the text, for $N=1024$ and 
two different initial connectivities ($K_{ini}=1.5$ and $K_{ini}=3.0$). 
Independent of the initial conditions chosen at random, the networks 
evolve to an average connectivity $K_{ev}=2.55 \pm 0.04$. The plot shows 
the time series and the corresponding cumulative means for $K_{ev}$. 
The evolutionary time $t$ is discrete, each time step representing a 
dynamical run on the evolved topology. Individual runs were limited to 
$T_{max}=1000$ iterations.}
\label{kevo_fig}       
\end{center}
\end{figure}
\begin{figure}[b]
\begin{center}
\includegraphics[scale=0.9]{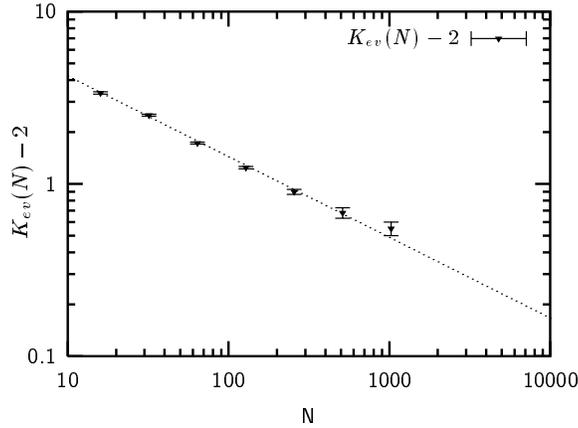}
\caption{The average connectivity of the evolved networks converges towards 
$K_c$ with a scaling law $\sim N^{-\delta}$, $\delta = 0.47 \pm 0.01$. 
For systems with $N \le 256$ the average was taken over $4 \cdot 10^6$ time 
steps, for $N=512$ and $N=1024$ over $5 \cdot 10^5$ and $2.5 \cdot 10^5$ 
time steps, respectively. Finite size effects from $T_{max}=1000$ may 
overestimate $K_{ev}$ for the largest network shown here.}
\label{kevo_scale_fig}      
\end{center}
\end{figure}
\begin{figure}[b]
\begin{center}
\includegraphics[scale=0.7]{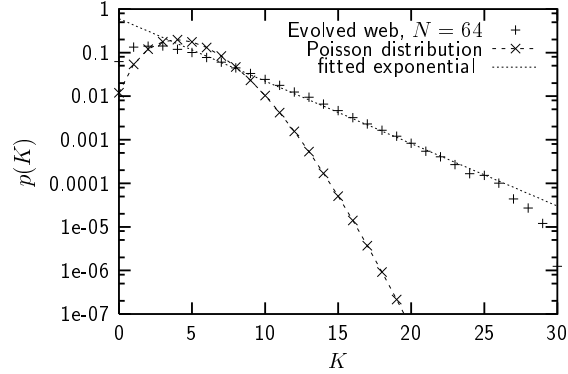}
\caption{Statistical distribution $p(K)$ of the number of inputs $K$ per 
node (gene) in the proposed model for a network of size $N=64$. Compared 
to the Poisson distribution for random networks with $\bar{K} = 4.46$, 
it shows a flatter decay $\propto \exp{[-K]}$.}
\label{evolvedinputdist_fig}      
\end{center}
\end{figure}
\begin{figure}[b]
\begin{center}
\includegraphics[scale=0.7]{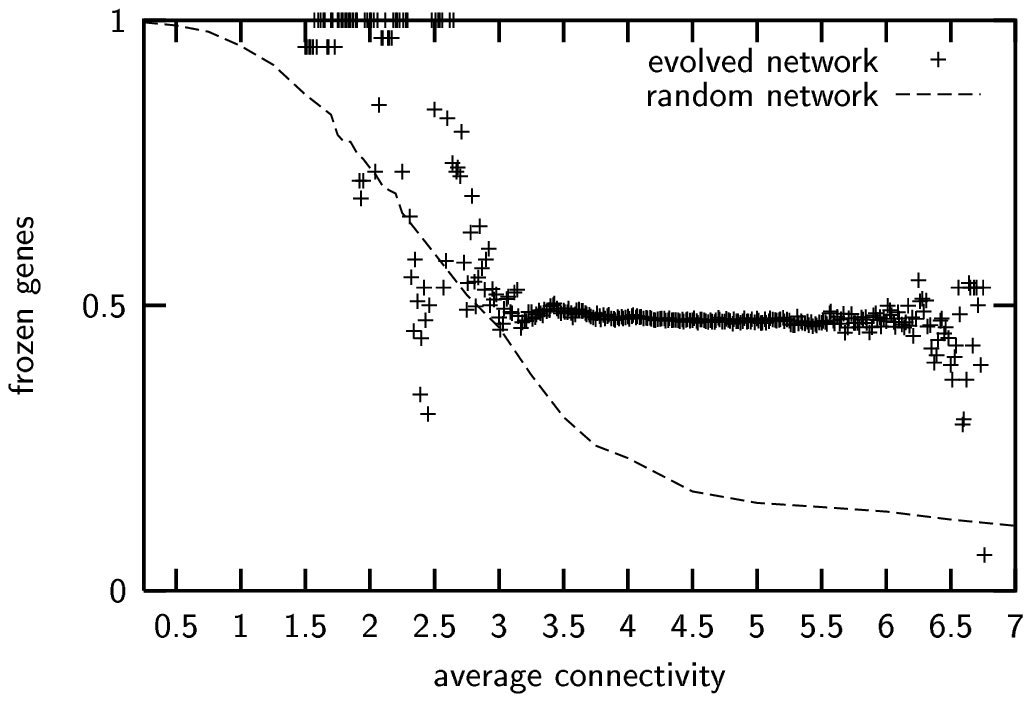}
\caption{The frozen component $C$ (fraction of frozen genes) as a function 
of the average connectivity for an evolved network of size $N=64$ (crosses). 
The dashed line shows the corresponding curve for random networks.}
\label{dynstep_fig}       
\end{center}
\end{figure}

\subsubsection{Results}
\label{subsubsec:actrewresults}
The typical picture arising from the model as defined above is shown in Fig.\ 
\ref{kevo_fig} for a system of size $N=1024$. Independent of the initial 
connectivity, the system evolves towards a statistically stationary state 
with an average connectivity $K_{ev}(N=1024)=2.55 \pm 0.04$. With varying 
system size we find that with increasing $N$ the average connectivity $\bar{K}$
approaches $K_c$ (which, for threshold $h=0$ as considered here, is found 
slightly below $\bar{K} = 2$ \cite{Rohlf2002}), see Fig.\ \ref{kevo_scale_fig}.
In particular, one can fit the scaling relationship 
\begin{eqnarray} 
K_{ev}(N) - 2 = c\cdot N^{-\delta} \label{kactrew_finitescale_eq}
\end{eqnarray}
to the measured connectivity values with $c = 12.4 \pm 0.5$ and $\delta = 
0.47 \pm 0.01$. In the evolutionary steady state, the average connectivity 
$\bar{K}$ of evolving networks exhibits limited fluctuations around the 
evolutionary mean $K_{evo}$ which are approximately Gaussian distributed, 
with a variance vanishing $\sim 1/N$ \cite{Rohlf2000}.
\begin{figure}[b]
\begin{center}
\includegraphics[scale=0.7]{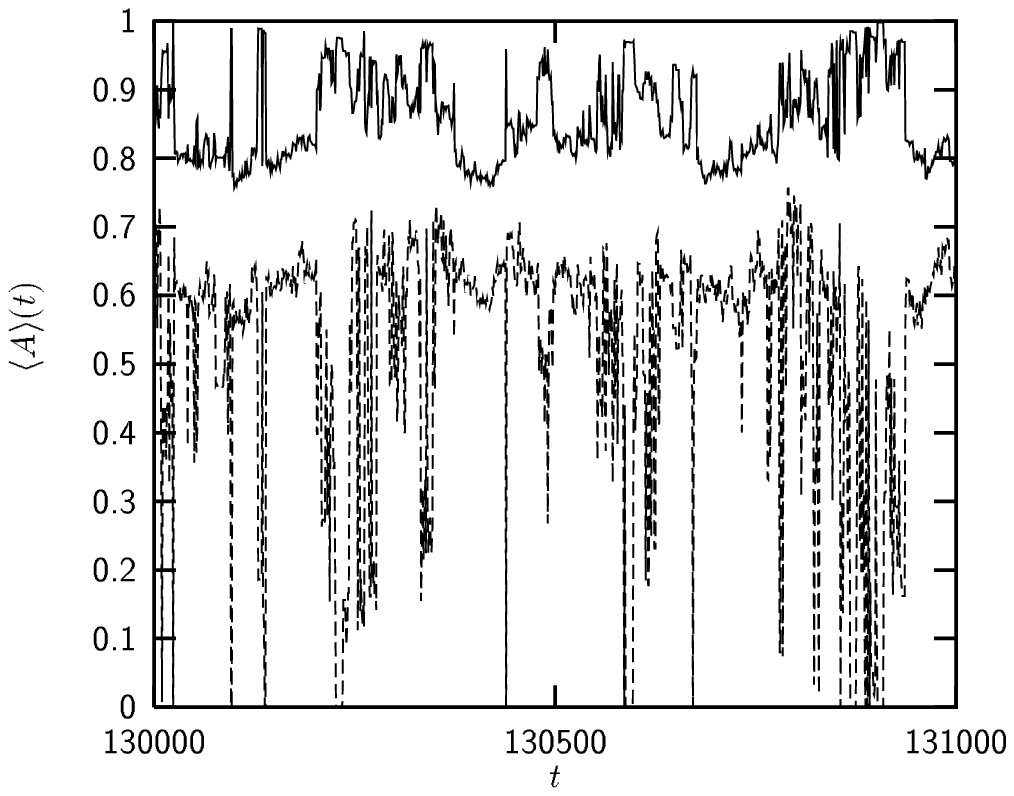}
\caption{Time series of the global average activity $\langle A\rangle$ 
(arbitrary time window of the evolutionary process). Upper curve: Signal 
averaged over the whole network, lower curve: Signal averaged over 
non-frozen nodes only.}
\label{acttimefig}       
\end{center}
\end{figure}
\begin{figure}[b]
\begin{center}
\includegraphics[scale=0.7,angle=270]{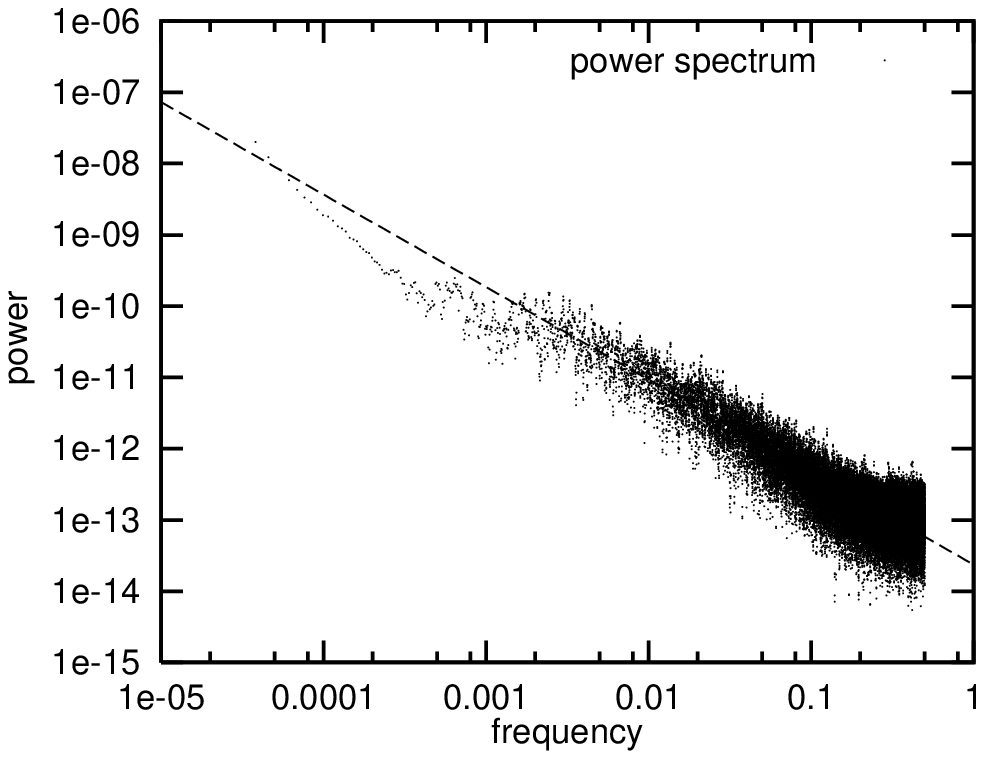}
\caption{Power spectrum of the global average activity $\langle A\rangle$ 
over $10^5$ evolutionary time steps, averaged over all network sites 
(compare upper curve in Fig. \ref{acttimefig}, double-logarithmic plot. 
The dashed line has slope $-1.298$.}
\label{powerspec_fig}     
\end{center}
\end{figure}

Going beyond averaged topological quantities, one can also measure the 
degree distributions of inputs and outputs in evolving networks, and 
compare it to what is expected for random networks (cf. section 
\ref{subsubsec:actprop}). In finite size networks, substantial deviations
from random graphs are found \cite{Rohlf2004a}: While the outdegree 
distribution stays close to the Poissonian of a random graph, evolved 
in-degree distributions are considerably flatter. For the averaged 
statistical distribution $p(K)$  of in-links (Fig.\
\ref{evolvedinputdist_fig}) of the evolving networks one observes a 
flat exponential decay
\begin{equation} 
p(K) \approx p_0\cdot \exp{[-\alpha K]}, 
\label{inputdist_eq}
\end{equation}
with $p_0, \alpha = \mbox{const}$. This observation indicates that the 
self-organized network state, at least for finite $N$, is substantially 
different from random networks with the same average connectivity. 
Since network evolution is based on co-evolutionary adaptation of 
dynamics and topology by local rewiring rules, this raises the question 
of whether the evolutionary statistically stationary state exhibits 
specific characteristics and correlations between dynamical and toplogical 
order parameters also on the {\em global scale}. This is indeed the case 
for finite $N$: If we compare, for example, the frozen component $C(\bar{K})$ 
or fraction of "frozen genes" for different values of connectivity 
fluctuations around the evolutionary mean $K_{evo}$ (Fig.\ \ref{dynstep_fig}), 
we observe that this curve exhibits a broad plateau where activity is 
stabilized at intermediate values, with almost step-like boundaries for 
small and large $\bar{K}$, whereas the corresponding curve for random 
networks is much smoother and decays earlier (compare also section 
\ref{subsubsec:actprop} for the phase transition observed in ensembles 
of random networks). This indicates that coevolution of dynamics and 
topology extends to a global scale, in spite of local rewiring events and 
a pronounced time scale separation between dynamical and topological updates 
\footnote{This time scale separation can be easily identified e.g.\ from 
step 4 of the adaptive algorithm summarized in Box 1: After $T$ dynamical 
system updates, one out of $N$ sites is rewired, hence time scale separation
is at the order of $T\cdot N$.}.

Last, let us characterize dynamics {\em on} the evolving networks, and 
investigate in how far it may exhibit signatures of self-organized 
critical behavior even in the finite size networks we studied so far 
(which, concerning average connectivity, are evidently super-critical). 
In contrast to random, noise-driven dynamics, where correlations decay 
fast (typically as an exponential), the self-organized critical state 
is characterized by non-trivial, long-range correlations in dynamical 
trajectories. A convenient measure to characterize such long-range 
correlations is the {\em power spectrum} of the dynamical time series.
Let us consider the autocorrelation function of a time signal $f(t)$, 
defined by
\begin{equation}
R(\tau)=\int_{-\infty}^{+\infty}f(t)f(t-\tau)\, dt.
\label{autocorr_eq}
\end{equation}
The power spectrum $G(f)$ is the Fourier transform of the autocorrelation
function, i.e.
\begin{equation}
G(f)=\int_{-\infty}^{+\infty}R(\tau)e^{-2\pi if\tau}d\tau.
\label{powspec_eq}
\end{equation}
In the case of time-discrete systems, the integrals are replaced by
the corresponding sums. For strongly (auto-)correlated systems, e.g.\ 
near the critical point, we typically expect a flat decay of the power
spectrum $G(f) \sim 1/f^{\alpha}$ with $\alpha\approx1$, while for a 
random walk, e.g., we would obtain $\alpha = 2$. The dynamical order 
parameter that we investigate is the {\em global average activity}
at evolutionary time step $t$:
\begin{equation}
\langle A\rangle(t)=\left|\frac{1}{N}\sum_{i=1}^{N}A_{i}(t)\right|.
\label{globact_eq}
\end{equation}
Figure \ref{acttimefig} shows a typical snapshot of the time series of 
$\langle A\rangle$ on evolving networks, the power spectrum is shown 
in Fig.\ \ref{powerspec_fig}. A least squares fit yields $G(f) 
\sim 1/f^{\alpha}$ with $\alpha = 1.298$ for the global average activity, 
i.e.\ a clear indication of long-range correlations in dynamics 
\cite{Rohlf2000}. Other measures of global dynamics also show evidence 
for criticality, for example, the statistical distribution of attractor 
periods is scale-free, as will be discussed in section 
\ref{subsec:liubassler} for a RBN variant of the model.

The self-organization towards criticality observed in this model is 
different from other known mechanisms exhibiting the amazingly 
general phenomenon of self-organized criticality (SOC) 
\cite{Bak1987,Bak1993,Paczuski1996,Sole1997}. 
Our model introduces a (new, and interestingly different) type of 
mechanism by which a system self-organizes towards criticality, 
here $K \rightarrow K_c$. This class of mechanisms lifts the notions 
of SOC to a new level. In particular, it exhibits considerable 
robustness against noise in the system. The main mechanism here is 
based on a topological phase transition in dynamical networks. 

In addition to the rewiring algorithm as described in this chapter, 
a number of different versions of the model were tested. Including the 
transient in the measurement of the average activity $A(i)$ results in 
a similar overall behavior (where we allowed a few time steps for the 
transient to decouple from initial conditions). Another version succeeds 
using the correlation between two sites instead of $A(i)$ as a mutation 
criterion (this rule could be called "anti-Hebbian" as in the context 
of neural network learning). In addition, this version was further changed  
allowing different locations of mutated links, both, between the tested 
sites or just at one of the nodes. All these different realizations exhibit 
the same basic behavior as found for the model above. Thus, the proposed 
mechanism exhibits considerable robustness. Interestingly, it has been shown 
that this mechanism leads to robust topological and dynamical self-organization
also in other classes of dynamical networks. In particular, Teuscher and 
Sanchez \cite{Teuscher2001} showed that this rule can be generalized to 
Turing neural networks and drives network evolution to $K_c =2$ in the 
limit of large $N$.

In the next subsection, we will discuss an extension of the model
that includes adaptation of thresholds in RTN, in addition to rewiring
of links. This extension still exhibits robust self-organization
as in the original model, however, exhibits several interesting new 
features, namely, symmetry breaking of evolutionary attractors,
and correlation of dynamical and toplogical diversity.
\begin{figure}[b]
\begin{center}
\includegraphics[scale=0.8]{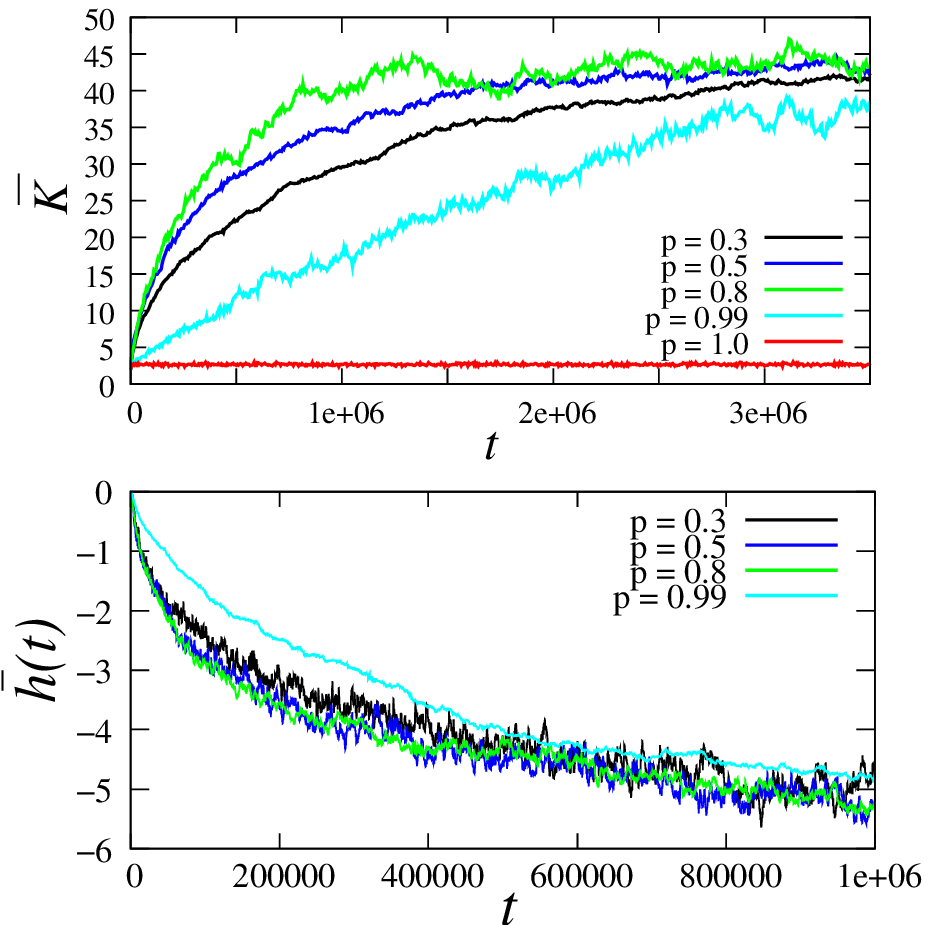}
\caption{{\em Upper panel:} Evolution of the average connectivity $\bar{K}$ 
of threshold networks, using the adaptive algorithm (cf. Fig. 1), for 
$N = 512$ and initial connectivity $\bar{K}_{ini} = 1$. Time series for 
five different values of $p$ are shown. {\em Lower panel:} The same for 
the average threshold $\bar{h}$.}
\label{khevo_fig}       
\end{center}
\end{figure}
\begin{figure}[b]
\begin{center}
\includegraphics[scale=0.9]{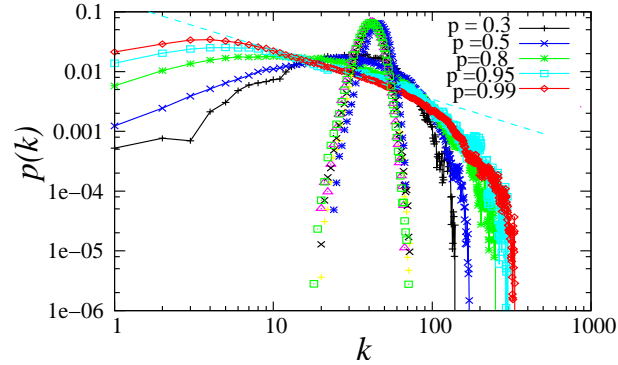}
\caption{{\em Line-pointed curves:} in-degree distributions of evolved 
networks, {\em data points only:} the corresponding out-degree distributions 
(($\triangle$) $p=0.3$, (+) $p =0.5$, (x) $p=0.8$, (*) $p =0.95$, 
(squares) $p=0.99$). Statistics was gathered over $10^6$ evolutionary steps, 
after a transient of $4\cdot 10^6$ steps. Networks had size $N = 512$. The 
dashed line has slope $-3/4$.}
\label{khinputdist_fig}      
\end{center}
\end{figure}
\begin{figure}[b]
\begin{center}
\includegraphics[scale=0.8]{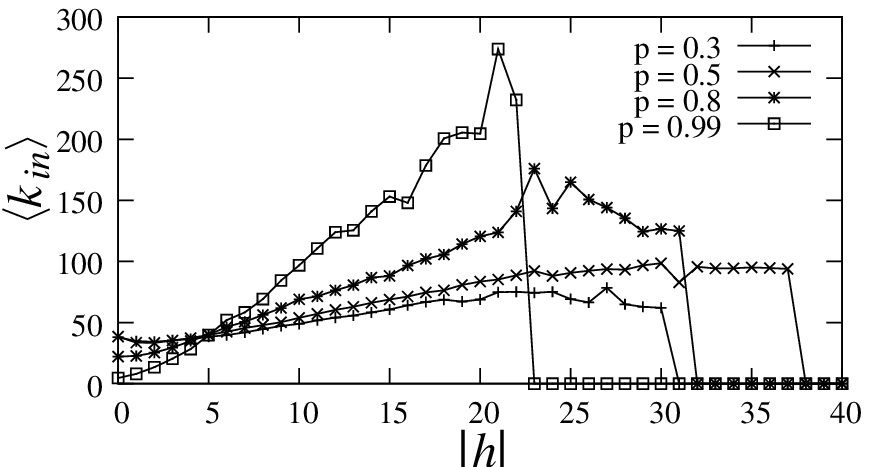}
\caption{Average number $\langle k_{in}\rangle$ of inputs for a given node 
in evolving networks, as a function of the respective nodes (absolute)
threshold $|h|$. Statistics was taken over $10^6$ rewiring steps, after a 
transient of $4\cdot 10^6$ steps. For all values $p < 1$, a clear positive 
correlation between $\bar{k}_{in}$ and $|h|$ is found.}
\label{khcorr_fig}      
\end{center}
\end{figure}

\subsection{Adaptive thresholds - time scale separation leads to 
complex topologies}
\label{subsec:adapth}
So far, we assumed that dynamical units in the networks are homogeneous 
(identical) with respect to their switching behavior, which for real 
world networks usually is a quite unrealistic assumption. Furthermore, 
recent studies have shown that inhomogeneity of thresholds leads to new
and unexpected phenomena in RTNs, e.g.\ an order-disorder transition 
induced by {\em correlations} between thresholds and input number of 
nodes \cite{Rohlf2007c}. In the general case of inhomogeneous thresholds, 
we have to modify Eq.\ \ref{rtnupdate_eq} such that
\begin{equation}
f_i(t) = \sum_{j=1}^N c_{ij}\sigma_j(t) + h_i,
\label{rtnsum_inhom_eq}
\end{equation}
where the indexed threshold $h_i$ now takes into account that thresholds 
can vary from node to node. The only restriction we impose is $h_i \le 0$, 
to make activation, i.e.\ $\sigma_i = +1$, more difficult.

We now introduce a minimal model linking regulation of activation 
thresholds and rewiring of network nodes in RTNs to local measurements 
of a dynamical order parameter \cite{Rohlf2007}. Adaptation of thresholds
opens up for the possibility of units that become heterogeneous with 
respect to their {\em dynamical} properties: Nodes with high thresholds 
are inert and switch their state only for few input configurations 
(similar to the effect of canalizing functions in RBNs), whereas nodes 
with low thresholds are more likely to switch. A new control parameter
$p \in [0,1]$ determines the probability of rewiring vs.\ threshold 
adaptations. In particular, the activity $A(i)$ of a site $i$ can be 
controlled in two ways: If $i$ is frozen, it can increase the probability 
to change its state by either increasing its number of inputs $k_i 
\rightarrow k_i + 1$, or by making its threshold  $h_i \le 0$ less 
negative, i.e.\ $|h_i| \rightarrow |h_i| - 1$. If $i$ is active, it 
can reduce its activity by adapting either $k_i \rightarrow k_i - 1$ 
or $|h_i| \rightarrow |h_i| + 1$. To realize this adaptive scheme, we 
have to modify step 4 in the adaptive algorithm of Box 1: 
{\bf A site $i$ is chosen at random and its average activity $A(i)$ 
during the last $T = \Gamma$ time steps is determined (in case no 
attractor is reached, $T=T_{max}/2$ is chosen). If $|A(i)| < 1$, 
then $k_i \rightarrow k_i-1$ with probability $p$ (removal of one 
randomly selected input). With probability $1-p$, adapt $|h_i| 
\rightarrow |h_i| + 1$ instead. If $|A(i)| = 1$, then $k_i 
\rightarrow k_i+1$ with probability $p$ (addition of a new input 
from a randomly selected site). With probability $1-p$, adapt $|h_i| 
\rightarrow |h_i| - 1$ instead. If $h_i = 0$, let its value unchanged.}
If the control parameter $p$ takes values $p > 1/2$, rewiring of nodes
is favored, whereas for $p < 1/2$ threshold adaptations are more likely.
Notice that the model discussed in the last subsection is contained as the 
limiting case $p = 1$ (rewiring only and $h_i = const. = 0$ for all sites).

{\em Results.} 
After a large number of adaptive cycles, networks self-organize into
a {\em global} evolutionary steady state. An example is shown in Fig.\ 
\ref{khevo_fig} for networks with $N =512$: starting from an initial value 
$\bar{K}_{ini} = 1$, the networks' average connectivity $\bar{K}$ first 
increases, and then saturates around a stationary mean value $\bar{K}_{evo}$; 
similar observations are made for the average threshold $\bar{h}$ (Fig.\ 
\ref{khevo_fig}, lower panel). The non-equilibrium nature of the system 
manifests itself in limited fluctuations of both $\bar{K}$ and $\bar{h}$ 
around $\bar{K}_{evo}$ and $\bar{h}_{evo}$. Regarding the dependence of 
$\bar{K}$ with respect to $p$, we make the interesting observation that 
it changes non-monotonically. Two cases can be distinguished: When $p = 1$, 
$\bar{K}$ stabilizes at a very sparse mean value $\bar{K}_{evo}$, e.g.\ 
for $N = 512$ at $\bar{K}_{evo} = 2.664 \pm 0.005$. When $p < 1$, the 
symmetry of this evolutionary steady state is broken. Now, $\bar{K}$ 
converges to a much higher mean value $\bar{K}_{evo} \approx 43.5 \pm 0.3$ 
(for $N=512$), however, the particular value which is finally reached is 
{\em independent of $p$}. On the other hand, {\em convergence times} 
$T_{con}$ needed to reach the steady state are strongly influenced by $p$: 
$T_{con}(p)$ diverges when $p$ approaches $1$ (compare Fig.\ 2 for $p = 0.99$). 
We conclude that $p$ determines the {\em adaptive time scale}. This is also 
reflected by the stationary in-degree distributions $p(k_{in})$ that vary 
considerably with $p$ (Fig. \ref{khinputdist_fig}); when $p \to 1$, these 
distributions become very broad. The numerical data suggest that a power law
\begin{equation}
\lim_{p \to 1} p(k_{in}) \propto k_{in}^{-\gamma}
\end{equation} 
with $\gamma \approx 3/4 \pm 0.03$ is approached in this limit (cf.\ Fig. 4, 
dashed line). At the same time, it is interesting to notice that the evolved 
out-degree distributions are much narrower and completely insensitive to $p$ 
(Fig. \ref{khinputdist_fig}, data points without lines). Hence, we make the 
interesting observation of a highly robust self-organization and homeostatic 
regulation of the average wiring density, while, at the same time in the limit 
$p \to 1$, time scale separation between frequent rewiring and rare threshold 
adaptation leads to emergence of complex, heterogeneous topologies, as 
reflected in the broad distribution of input numbers approaching a power law.
Obviously, we have a non-trivial coevolutionary dynamics in the limit $p \to 1$ 
which is significantly different from the limit of small $p$. This is also 
indicated by the emergence of strong correlations between input number and 
thresholds in this limit (see the steep increase of the curves for $p > 0.5$ 
in Fig.\ \ref{khcorr_fig}), while in the limit of small $p$ correlations are 
weak.

To summarize, we find that coevolution of both rewiring and threshold 
adaptation with the dynamical activity on RTNs leads to a number of 
interesting new effects: We find spontaneous symmetry breaking into a new 
class of adaptive networks that is characterized by increased heterogeneity 
in wiring topologies and emergence of correlations between thresholds and 
input numbers. At the same time, we find a highly robust regulation of the 
average wiring density which is independent of $p$ for any $p < 1$.

In the next subsection, we will discuss another generalization of the 
adaptive, coevolutionary scheme of activity-dependent rewiring to Random 
Boolean Networks, which was introduced by Liu and Basler \cite{Liu2006}.

\subsection{Extension to Random Boolean Networks}\label{subsec:liubassler}
Activity-dependent rewiring was originally introduced for Random Threshold 
Networks, as discussed in section \ref{subsubsec:actrewmodel},
the basic adaptive scheme, however, can be generalized to other
classes of dynamical systems. Since RTN are a subclass of Random Boolean
Networks, one possible direction of generalization is to apply this 
coevolutionary, adaptive rule to RBNs. Compared to RTNs, rewiring by local 
dynamical rules in RBN comes with an additional complication: While in RTNs
the dynamical transition rule is the same for all network sites (the 
evaluation of the weighted sum of regulatory inputs, cf.\ Eq.\ \ref{rtnsum_eq}),
switching of network nodes in RBNs is governed by individual logical functions 
that vary from node to node {\em and} depend on the input number $k$. 
If, for example, we have a node with two inputs and a logical AND function 
of these two inputs assigned (compare the example in Fig. \ref{RBNscheme}),
there does not exist a well-defined mapping that would assign a new logical
function to this node in the case we change its input number to $k=1$ or $k=3$.

Liu and Bassler \cite{Liu2006} suggested two variants of activity-dependent
rewiring to overcome the problem associated to the reassignment of
logical functions: in the first variant, only the node that is rewired
at evolutionary time step $t$ is assigned a new logical function which 
is randomly drawn out of the $2^{2^k}$ possible Bollean functions of $k$
inputs (where $k$ is the new input number {\em after} rewiring). 
The adaptive algorithm that was applied in this study is summarized
in Box 2.
\begin{svgraybox}
{\bf Box 2: Adaptive algorithm for activity-dependent rewiring in RBN}\\
\begin{enumerate}
\item Start with a homogeneous RBN, $G(N,K_0)$ with uniform in-degree
      connectivity $K_i=K_0$ for all $N$, and generate a random Boolean 
      function $f_i$ for each node $i$.
\item Choose a random initial system state $\Sigma(0)$. Update the state 
      using Eq.~\ref{statevec_eq} and find the dynamical attractor. 
\item Choose a node $i$ at random and determine its average
      activity $A(i)$ over the attractor.
\item Change the network topology by rewiring the connections to the
      node chosen in the previous step. If it is frozen, then a new 
      incoming link from a randomly selected node $j$ is added to it. 
      If it is active, then one of its existing links is randomly selected 
      and removed.
\item The Boolean functions of network are regenerated. Two different 
      methods have been used:
\begin{itemize}
      \item Annealed model: A new Boolean function is generated for every
            node of the network.
      \item Quenched model: A new Boolean function is generated only for the 
            chosen node $i$, while the others remain what they were previously.
\end{itemize}
\item Return to step 2.
\end{enumerate}
\end{svgraybox}
\begin{figure}[htb]
\begin{center}
\includegraphics[scale=0.3]{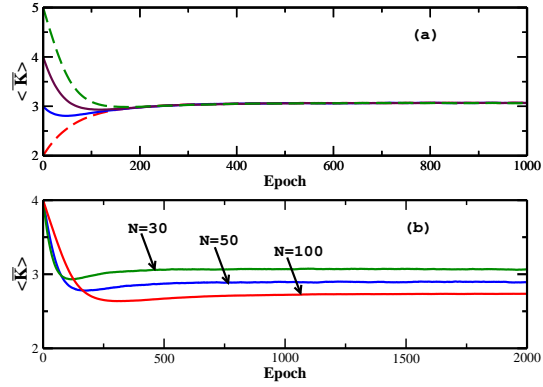}
\caption{(a). Evolution of the ensemble averaged in-degree connectivity in 
the annealed model, as studied by Liu and Bassler \cite{Liu2006}, for 
networks of size $N=30$. The networks in each ensemble initially start 
from different uniform connectivity, $K_0 = 2, 3, 4$, and 5, but reach 
a same statistical steady state $\langle \overline{K}\rangle=3.06$. 
(b). Evolution of ensemble averaged in-degree connectivity for networks 
of three different size $N=30,50$, and 100 in the annealed model.}
\label{AverageK}
\end{center}
\end{figure}
\begin{figure}[b]
\begin{center}
\includegraphics[scale=0.25]{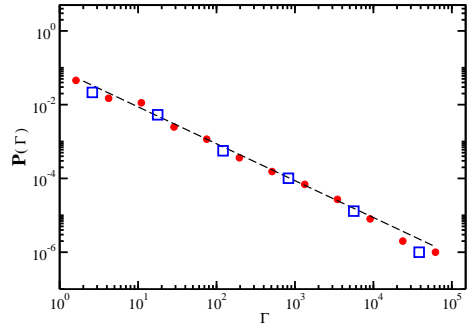}
\caption{Power law distribution of steady state attractor period $\Gamma$ 
in both annealed (circle) and quenched (square) models as studied by Liu 
and Bassler \cite{Liu2006}, for $N=200$ systems. The dashed straight line 
has a slope of $1.0$.}
\label{attr_dist_BLfig}      
\end{center}
\end{figure}
For simplicity, all random Boolean functions are generated with $p=1/2$, 
and therefore all Boolean functions with the same in-degree are equally 
likely to be generated. 

{\em Results.} Liu and Bassler show that for both variants of the model,
robust self-organization of network topology is found. Independent from 
the initial network realization, network evolution always converges to a 
characteristic average connectivity $K_{ev}(N)$. Graph (a) of Fig.\ 
\ref{AverageK} shows the evolution of the average in-degree connectivity 
$\overline{K}$ for networks of size $N=30$ in the annealed variant of the 
model, with results obtained by beginning with networks with different 
uniform connectivity $K_0=2$, 3, 4, and 5. Each curve is the average 
of 15,000 independent realizations of the network evolution. All curves
approach the same final statistical steady state that has an average 
in-degree connectivity $\langle \overline{K}\rangle =3.06$. The steady 
state value of $\langle \overline{K} \rangle$ depends on the size of the 
system as shown in graph (b) of Fig.~\ref{AverageK}. Starting with networks 
that all have the same initial uniform connectivity $K_0=4$, but which have 
different size $N=30,50$, and 100, one finds that larger networks evolve to 
steady states with smaller values of $\langle \overline{K} \rangle$. 

Given the steady state value $\langle\overline{K} \rangle=2$ in the large 
network limit $N\rightarrow\infty$, Liu and Bassler also studied the 
finite-size effects in the model. They found that the values of $\langle 
\overline{K}(N)\rangle$ for finite $N$ obey the scaling function
\begin{equation}
\langle\overline{K}(N)\rangle - 2 = c\, N^{-\delta}.
\label{kboolrew_eq}
\end{equation}
Fitting the data to this function, we find that the coefficient is $c=2.50 
\pm 0.06$ and the exponent is $\delta=0.264\pm 0.005$. Thus the value of
$\langle\overline{K}(N)\rangle$ is always larger than 2 for finite $N$. 
Note that steady state values of the average connectivity in RTNs have a 
similar scaling form, but with slightly different values of the scaling 
parameters (cf.\ section \ref{subsubsec:actrewresults}, Eq.\ 
\ref{kactrew_finitescale_eq}).

In order to probe the dynamical nature of evolved steady states the 
authors computed the distribution $P(\Gamma)$ of steady state attractor
period $\Gamma$ in the ensemble of RBNs simulated. The distribution
has a broad, power-law behavior for both the annealed and quenched
variants of the model. Figure~\ref{attr_dist_BLfig} shows the results 
for networks with $N=200$. A power-law distribution of attractor periods 
is a typical signature of critical dynamics, hence, similar to the 
results discussed in section \ref{subsubsec:actrewresults} for the 
self-organization of the global average activity, this finding indicates 
that {\em dynamics} exhibits close-to critical behavior already for 
finite size networks, while {\em toplogical} criticality is attained 
in the limit of large $N$.

To summarize, the results of this study give strong evidence that the local, 
adaptive coevolutionary principle {\em frozen nodes grow links, active nodes 
lose links} leads to robust self-organization not only in RTNs, but also in 
the more general class of RBNs, and hence has the potential to be generalized 
to large classes of dynamical systems.

\subsection{Correlation-based rewiring in neural networks}
\label{subsec:correlrew}
In this section, we will review a different adaptive coeveolutionary scheme 
of network self-organization which is based on the basic paradigm 
{\em correlated activity connects, decorrelated activity disconnects.}  
This local, topology-evolving rule is inspired by the idea of {\em Hebbian 
learning} in neural networks \cite{Hebb1949}, and, consequently, was 
studied first for discrete neural networks with architectural and 
dynamical constraints motivated by corresponding observations in the 
brain \cite{Bornholdt2003}. 
In particular, an explicit parameterization of space on a 
two-dimensional grid is given, and dynamics is not deterministic any 
more, in contrast to the models discussed in the previous sections. 
The core result of this study is that network self-organization by 
correlation-driven rewiring is robust even when spatial constraints 
are present and dynamics is affected by noise. 
Correlation-driven rewiring can be considered as a natural extension 
of the basic, activity-driven rewiring, as it exploits long range 
correlations naturally emerging near phase transitions, and thereby is 
particularly suited for neural networks where information processing 
takes place typically in form of correlated activity.
Let us now briefly motivate the correlation based self-organization 
in the context of neural networks.

Neural networks with asymmetric connectivity patterns often exhibit regimes 
of chaotic dynamics \cite{Molgedey1992}. In networks whose central 
function is information transfer, these regimes would instantly render 
them useless. Consider, for example, model neural networks with asymmetric
synaptic couplings, where a percolation transition between regimes of 
ordered and disordered dynamics is known \cite{Kuerten1988a}. In the 
disordered phase, which occurs for densely connected networks, already 
small perturbations percolate through the networks.\footnote{This is 
reminiscent of avalanche-like propagation of activity in the brain which is 
observed in some diseases of the central nervous system \cite{Schroeder1998}.}
In such networks, developmental processes that change connectivity always 
face the risk of driving the network into the highly connected regime 
(where chaotic dynamics prevails), as long as no explicit mechanism is 
given that controls the global degree of connectivity. 

In a correlation-based rewiring rule we will exploit that also the 
average correlation between the activities of two neurons contains 
information about global order parameter of network dynamics. 
The network can then use this approximate order parameter to guide 
the developmental rule. A possible adaptive scheme is that new 
synaptic connections preferentially grow between correlated neurons, 
as suggested by the early ideas of Hebb \cite{Hebb1949} 
and the observation of activity-dependent neural development 
\cite{Trachtenberg2002,Ming2002,Engert1999,Ooyen2001}.  
In the remainder of this section let us recapitulate this problem 
in the framework of a specific toy model \cite{Bornholdt2003}. 
First a neural network model with a simple mechanism of synaptic 
development is defined. Then, the interplay of dynamics on the 
network with dynamics of the network topology is modeled.
Finally, robustness of self-organizing processes in this model
and possible implications for biological systems are discussed.
 
\subsubsection{Model}
\label{subsubsec:correlmodel}
Let us consider a two-dimensional neural network with random asymmetric 
weights on the lattice. The neighborhood of each neuron 
is chosen as its Moore neighborhood with eight neighbors.\footnote{The 
choice of the type of neighborhood is not critical, however, here
the Moore neighborhood is more convenient than the von Neumann 
type since, in the latter case, the critical link density 
(fraction of nonzero weights) at the percolation threshold 
accidentally coincides with the attractor of the trivial 
developmental rule of producing a link with $p=0.5$. In general, 
also random sparse neighborhoods would work as demonstrated in 
\cite{Bornholdt2000}.} The weights $c_{ij}$ are randomly drawn from a 
uniform distribution $c_{ij} \in \left[-1, +1 \right]$ and are nonzero
between neighbors, only. Note that weights $c_{ij}$ are asymmetric, i.e., 
in general, $c_{ij} \not= c_{ji}$. Within the neighborhood of a node, 
a fraction of its weights $c_{ij}$ may be set to $0$. The network consists 
of $N$ neurons with states $\sigma_i = \pm 1$ which are updated in parallel 
with a stochastic Little dynamics on the basis of inputs received from the 
neighbor neurons at the previous time step:
\begin{eqnarray}
\sigma_i(t+1)=
\begin{cases}
+1 \quad \mbox{with probability } {g_\beta}\left(f_i(t)\right) \\
-1 \quad \mbox{with probability } 1-{g_\beta}\left(f_i(t)\right)
\end{cases}
\label{littledyn_eq}
\end{eqnarray}
where
\begin{eqnarray}
g_\beta( f_i(t)) = \frac{1}{1 +  e^{- 2 \beta f_i(t)}}
\label{transfer_eq}
\end{eqnarray}
with the inverse temperature $\beta$. The transfer function $f_i(t)$ 
is evaluated according to Eq. \ref{rtnsum_inhom_eq}, that defines
dynamics of threshold units with individually assigned thresholds.
The threshold is chosen here as $h_i= -0.1+\gamma$ and includes a small 
random noise term $\gamma$ from a Gaussian of width $\epsilon$. 
This noise term is motivated by the slow fluctuations observed in 
biological neural systems \cite{Abraham2001,Abraham1997}. 

The second part of the model is a slow change of the topology of the
network by local rewiring of synaptic weights: If the activity of two 
neighbor neurons is on average highly correlated (or anticorrelated), they 
will obtain a common link. If their activity on average is less correlated, 
they will lose their common link. The degree of correlation in the dynamics 
of pairs of nodes is quantified by the {\em average correlation}, as 
defined in Eq. \ref{corrdef_eq} in section \ref{subsubsec:actdef}. The 
full model dynamics is then realized by the algorithm summarized in Box 3.
\begin{svgraybox}
{\bf Box 3: Adaptive algorithm for correlation-dependent rewiring in 
neural networks}\\
\begin{enumerate}
\item Start with a random network with an average connectivity (number of 
      nonzero weights per neuron) $K_{ini}$ and a random initial state 
      vector $\Sigma(0) = (\sigma_1(0),...,\sigma_N(0))$.
\item For each neuron $i$, choose a random threshold $h_i$ from a Gaussian 
      distribution of width $\epsilon$ and mean $\mu$.
\item Starting from the initial state, calculate the new system state 
      applying Eq. \ref{littledyn_eq} using parallel update. Iterate this 
      for $\tau$ time steps.
\item Randomly choose one neuron $i$ and one of its neighbors $j$ and 
      determine the average correlation according to Eq. \ref{actdef_eq} 
      over the last $\tau/2$ time steps. (Alternatively, the correlation 
      can be obtained from a synaptic variable providing a moving average 
      at any given time).
\item If $|\mbox{Corr}(i,j)|$ is larger than a given threshold $\alpha$, 
      $i$ receives a new link $c_{ij}$ from site $j$ with a weight chosen 
      randomly from the interval $c_{ij} \in \left[-1,1\right]$.\footnote{
      Also binary weights could be used as in Ref.\ \cite{Bornholdt2000}.}
      If $|\mbox{Corr}(i,j)| \leq \alpha$, the link $c_{ij}$ is set to $0$ 
      (if nonzero).  
\item Go to step 2 and iterate, using the current state of the network as 
      new initial state.
\end{enumerate}
\end{svgraybox}
\noindent
The dynamics of this network is continuous in time, with neuron update 
on a fast time scale and topology update of the weights on a well-separated 
slow "synaptic plasticity" time scale. Note that the topology-changing 
rule does not involve any global knowledge, e.g., about attractors.
\begin{figure}[b]
\begin{center}
\includegraphics[scale=0.8]{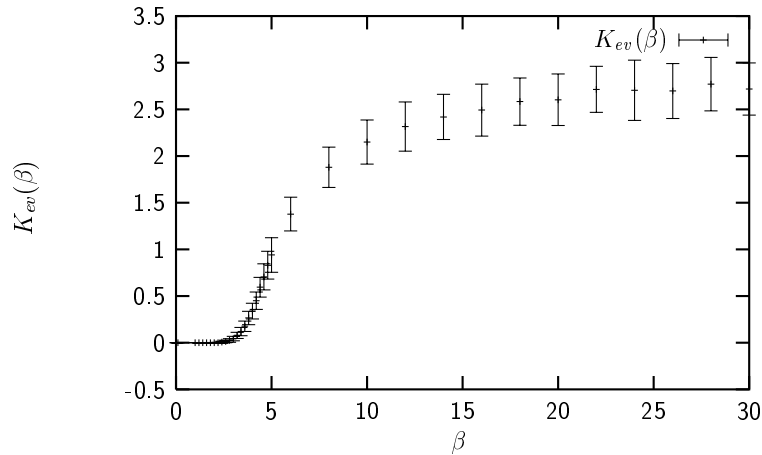}
\caption{Evolved average connectivity $K_{ev}$ as a function of the inverse 
temperature $\beta$. Each point is averaged over $10^5$ time steps in a 
network of size $N=64$ and $\alpha=0.5$. After Bornholdt and R\"ohl 
\cite{Bornholdt2003}.}
\label{fig3_neuro}    
\end{center}
\end{figure}
\begin{figure}[b]
\begin{center}
\includegraphics[scale=0.8]{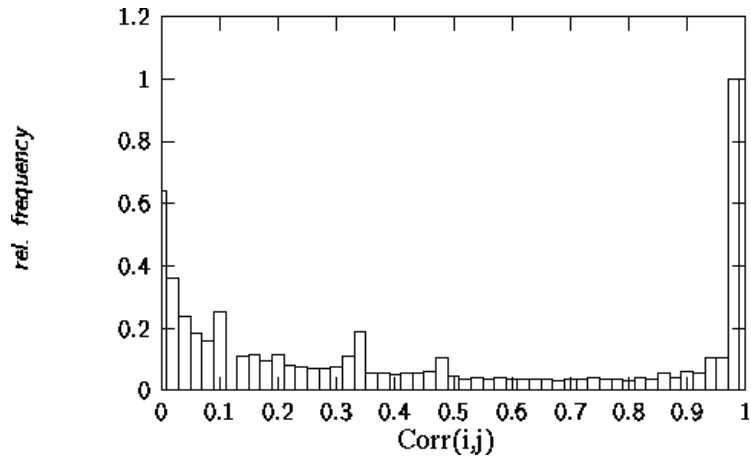}
\caption{Histogram of $\mbox{Corr}(i,j)$ for a network evolving in time, with
$N=64$ and $\beta=10$, taken over a run of $4\times10^5$ time steps, 
according to the model of Bornholdt and R\"ohl \cite{Bornholdt2003}.}
\label{fig5_neuro}       
\end{center}
\end{figure}

\subsubsection{Results}
\label{subsubsec:correlresults}
Independent of the initial conditions the networks evolve to a specific 
average connectivity. Parameters are $\beta = 25$, $\epsilon=0.1$, 
a correlation cutoff $\alpha = 0.8$, and an averaging time window of 
$\tau=200$. One observes that the continuous network dynamics, including 
the slow local change of the topology, results in a convergence of the
average connectivity of the network to a characteristic value which is
independent of initial conditions.

Finite size scaling of the resulting average connectivity indicates
the convergence towards a characteristic value for large network size 
$N$ and exhibits the scaling relationship
\begin{eqnarray}
K_{ev}(N) = a N^{-\delta} + b 
\label{kcorrelrew_finite_eq}
\end{eqnarray}
with $a =1.2\pm0.4$, $\delta =0.86\pm0.07$, and $b=2.24\pm0.03$. Thus, 
in the large system size limit $N \rightarrow \infty$ the networks 
evolve towards $K_{ev}^\infty = 2.24\pm0.03$. The self-organization 
towards a specific average connectivity is largely insensitive to 
thermal noise of the network dynamics, up to $\approx 10 \%$ of thermal 
switching errors (or $\beta >10$) of the neurons. This indicates that 
the structure of a given dynamical attractor is robust against a large 
degree of noise. Figure \ref{fig3_neuro} shows the evolved average 
connectivity as a function of the inverse temperature $\beta$.

While the stability of dynamical attractors on an intermediate time
scale is an important requirement for the local sampling of neural
correlation, on the long time scale of global topological changes,
switching between attractors is necessary to ensure ergodicity at 
the attractor sampling level. The second source of noise, the slow
random change in neural thresholds as defined in step (2) of the
algorithm, is closely related to such transitions between attractors.
While, in general, the model converges also when choosing some arbitrary 
fixed threshold $h$ and omitting step (2) from the algorithm, a small 
threshold noise facilitates transitions between limit cycle attractors 
\cite{McGuire2002} and thus improves sampling over all attractors of 
a network, resulting in an overall increased speed and robustness of 
the convergence. An asynchronous change of the threshold $h_i$, updating 
one random $h_i$ after completing one sweep (time step) of the network, 
leads to similar results as the parallel rule defined above.

The basic mechanism of the observed self-organization in this system is
the weak coupling of topological change to an order parameter of the 
global dynamical state of the network, and thus is different from the
mechanism of extremal dynamics, underlying many prominent models of 
self-organized criticality \cite{Bak1987,Bak1993}. To illustrate this,
let us for a moment consider the absolute average correlation
$|\mbox{Corr}(i,j)|$ of two neurons which is the parameter used as a
criterion for the rewiring process. It can be shown that this quantity 
undergoes a phase transition depending on the average connectivity 
$\bar{K}$ which is similar to the transition of the frozen component 
observed in RTN (Fig.\ref{frozcomp_fig}, cf. section \ref{subsubsec:actprop}).
Note that the correlation is large for networks with small connectivity,
and small for networks that are densely connected. The rewiring rule 
balances between these two regimes: For high correlation, it is more 
likely that a link is created, at low correlation, links are vanishing. 
The balance is reached most likely in the region of the curve where the 
slope reaches its maximum, as here the observed correlation reacts most 
sensitively to connectivity changes. As the steep portion of the 
correlation curve occurs in a region of small connectivities where also 
the critical connectivity $K_c\approx 2$ of the network is located, 
this makes the correlation measure sensitive to the global dynamical 
state of the network and potentially useful as an approximation of 
the order parameter. Synaptic development dependent on averaged 
correlation between neurons can thus obtain approximate information 
about the global dynamical state of the network as is realized in the 
above toy model with a simple implementation on the basis of a threshold 
$\alpha$. The exact choice of the threshold $\alpha$ is not critical, 
which can be seen from the histogram of the absolute correlation shown 
in Fig.\ \ref{fig5_neuro} for a typical run of the model.
Correlations appear to cluster near high and near low values such
that the cutoff can be placed anywhere inbetween the two regimes.
Even a threshold value close to $1$, as compared with the correlation
cutoff $\alpha=0.8$ used in the simulations here, only leads to a
minor shift in $K_{ev}$ and does not change the overall behavior.

Up to now we focused on changes of the network structure as a result of
the dynamics on the network. A further aspect is how the structural
changes affect the dynamics on the network itself. Do also dynamical
observables of the networks self-organize as a result of the observed
convergence of the network structure? An interesting quantity in this
respect is the average length of periodic attractors.

Indeed, this dynamical observable of the network dynamics converges
to a specific value independent of the initial network, similarly to
the convergence of the structural parameter $\bar{K}$ considered earlier.
From the $\bar{K}$ dependency of the neural pair correlation we have seen
above that the rewiring criterion tends to favor connectivities near
the critical connectivity of the network. Does also the evolved average 
attractor length relate to critical properties of the percolation transition? 
An approximate measure of this aspect is the finite size scaling of the 
evolved average period.

For static networks we find that the attractor lengths typically scale 
exponentially with $N$ in the overcritical regime, but less than linearly
in the ordered regime. For the evolved connectivity $K_{ev}$ in our model, 
we observe scaling close to criticality. Large evolved networks exhibit 
relatively short attractors, which otherwise for random networks in the 
overcritical regime could only be achieved by fine tuning. The self-organizing 
model studied here evolves non-chaotic networks without the need for parameter 
tuning.

In a continuously running network, robust self-organization of the
network towards the percolation transition between ordered and
disordered dynamics is observed, independent of initial conditions
and robust against thermal noise. The basic model is robust against
changes in the details of the algorithm. We conclude that a weak
coupling of the rewiring process to an approximate measurement of
an order parameter of the global dynamics is sufficient for a robust
self-organization towards criticality. In particular, the order parameter 
has been estimated solely from information available on the single 
synapse level via time averaging of correlated neural activities.

\section{Summary and Outlook}
\label{sec:summary}
We reviewed models of topological network self-organization by 
{\em local dynamical rules}. Two paradigms of local co-evolutionary 
adaptation were applied to discrete dynamical networks: The principle 
of {\em activity-dependent rewiring} (active nodes lose links, frozen 
nodes aquire new links), and the principle of {\em correlation-dependent 
rewiring} (nodes with correlated activity connect, decorrelated nodes 
disconnect). Both principles lead to robust self-organization of {\em global} 
network topology and -dynamics, without need for parameter tuning. Adaptive 
networks are strikingly different from random networks: they evolve 
inhomogeneous topologies and broad plateaus of homeostatic regulation, 
dynamical activity exhibits $1/f$ noise and attractor periods obey a 
scale-free distribution. The proposed co-evolutionary mechanism of 
topological self-organization is robust against noise and does not 
depend on the details of dynamical transition rules. Using finite-size 
scaling, it was shown that networks converge to a self-organized 
critical state in the thermodynamic limit. 

The proposed mechanisms of coevolutionary adaptation in dynamical networks 
are very robust against changes in details of the local rewiring rules - 
in particular, they only require a local estimate of some dynamical order 
parameter in order to achieve network adaptation to criticality. 

A classical route to self-organized criticality is the feedback
of an order parameter onto local dynamics of a system \cite{Sornette1992}. 
The local rewiring rules considered here extend this idea to using 
only approximate local estimates of a global order parameter. 
As seen in the examples above, locally measured averages prove to be 
sufficient for self-organized criticality. In particular, in the 
presence of a time-scale separation between fast dynamics on the 
networks and slow topology evolution, the evolutionary steady state 
naturally provides a quasi-ergodic sampling of the phase space near
the critical state, such that accurate order parameter values are not necessary. 
In the models considered in this review, an estimate of the order parameter 
is achieved by local averaging over the switching activity of single nodes, 
or over the dynamical correlation of pairs of nodes. 

A number of open questions remains to be addressed in the context of 
the class of adaptive networks considered here. 
The robust network self-organization observed in the models, approaching  
criticality in the limit of large $N$, is far from being understood in detail.  
With regard to dynamics, evolving networks exhibit pronounced differences to 
random networks with comparable connectivity. In particular, the frozen 
component exhibits for finite $N$ a plateau around the evolutionary mean 
$K_{evo}$, with step-like discontinuities when the average wiring density 
substantially departs from $K_{evo}$ (Fig. \ref{dynstep_fig}); for comparison, 
the corresponding curves for random networks show a smooth decay. 
This may suggest that the self-organized critical state rather exhibits 
characteristics of a first order phase transition, while order parameters 
in random discrete dynamical networks typically exhibit second order 
transitions at $K_c$. With regard to topology, deviations from random 
networks become particularly pronounced when dynamical units are allowed 
to diversify with respect to their switching behavior during evolution, 
leading to symmetry breaking and emergence of scale-free in-degree distributions; 
again, these observations are hard to explain in the context of the 
traditional statistical mechanics approach based on random ensembles 
of networks. 

In the studies reviewed in this chapter, evolving networks were treated 
as completely autonomous systems, without coupling to an external 
environment. An important step in future research will be to introduce 
network-environment interaction and study network evolution under the 
influence of external signals or perturbations; this also connects to 
the particularities of information processing in self-organized
critical networks, and the idea of optimal adaptation at the 
"edge of chaos"\cite{Bertschinger2005}. 

Finally, let us widen the scope of these models beyond their theoretical 
value, and discuss possible applications.

On the one hand, they represent prototypical models of self-organized 
critical dynamical networks, toy models that demonstrate possible  
mechanisms for dynamical networks to adapt to criticality. 
On the other hand, these mechanisms are not limited to the extremely 
simple toy models discussed here, and may themselves occur in 
natural systems. They do not depend on fine-tuning of parameters or 
details of the implementation, and they are robust against noise. 
This is contrary to standard mechanisms of self-organized criticality 
\cite{Bak1987,Bak1993} which are sensitive to noise \cite{Schmoltzi1995}
and, therefore, not easily applied to natural systems. 
Network self-organization as reviewed above, however, is itself 
defined on the basis of stochastic dynamical operations (update of 
randomly selected links, noisy local measurement of order parameter, 
for example). We therefore expect that these mechanisms can occur 
in natural systems quite easily. 

A strong need for adaptive mechanisms is present in nervous systems, 
where assemblies of neurons need to self-adjust their activity 
levels, as well as their connectivity structure 
\cite{Trachtenberg2002,Beggs2003,Stewart2008,Gireesh2008}.
The mechanism discussed here is one possible route to adaptivity 
in natural neural networks. It can serve as the basis for more 
biologically detailed models 
\cite{Bertschinger2005,Beggs2008,Levina2008}. 

Further applications of the network adaptation models are conceivable, 
e.g.\ to socio-economic systems. Network adaptation could in principle 
occur in adapting social links or economic ties of humans acting as 
agents in complex social or economic systems.

\bibliographystyle{spmpsci} 

\end{document}